# Business Process Modeling:

## Blueprinting

Sabah Al-Fedaghi
Computer Engineering Department
Kuwait University
Kuwait
sabah.alfedaghi@ku.edu.kw

*Abstract*—**This paper presents a flow-based methodology for capturing processes specified in business process modeling. The proposed methodology is demonstrated through re-modeling of an IBM Blueworks case study. While the Blueworks approach offers a well-proven tool in the field, this should not discourage workers from exploring other ways of thinking about effectively capturing processes. The diagrammatic representation presented here demonstrates a viable methodology in this context. It is hoped this explicit analysis of diverse fundamental approaches will benefit the research in the field and also advance current practices.**

## I. INTRODUCTION

Current tendencies in development of information and communication systems are leaning toward higher levels of abstraction in process modeling. *Process modeling* aims at capturing the reality of processes [1] through diagrams that symbolically depict the capture, manipulation, and handling of data and information between a system and its environment and among system components. Business process modeling is an example of such a process. The term "business" is interchangeable with "organization," a term applicable to all sorts of organizations such as government agencies and departments, charities, mutuals and cooperatives, etc. [2]. The term Business Process Model refers to a diagram that describes flows and activities in a particular business or organizational unit [2].

Process modeling takes diverse approaches found in the literature as well as in practice; e.g., software engineering, enterprise modeling, knowledge modeling, simulation and quantitative analyses, and workflow systems. Without loss of generality, this paper focuses on process modeling as adopted in *IBM Blueworks* [3-4], which captures business process blueprints and creates BPMN diagrams. Here the methodology is to start the "blueprinting process" [3] by generating *milestones*, followed by identifying *activities* in "discovery sessions" with the process owner and subject matter experts. Identifying milestones reflects a thinking paradigm, a certain way of generating elements, and directs attention to identifying interactions in a manner that affects how we *link* concepts.

The most noteworthy aspect of the new Blueworks Live offering extends it beyond being a platform for process discovery and modelling – and into a platform for process automation and execution. Blueworks … focused on helping tackle the many dozens or hundreds of lightweight processes that are not properly with IT today. [6]

According to Peisl [6],

More and more customers are using IBM Blueworks Live for business process discovery and modeling, as well as for simplified analysis and … people LOVE to model their business processes with this tool, which is hosted in the cloud / managed by IBM and does not require any IT support ...

This paper investigates the underlying theoretical principle at the base of the "blueprinting process" and proposes a flow-based methodology of thinking to apply to capturing processes in business process modeling. The proposed methodology is demonstrated through re-modeling of a Blueworks case study. This explicit analysis of diverse basic approaches should benefit the research in this field and could also advance current proven practices such as Blueworks.

The next section introduces a flow-based diagrammatic language to be used both as a thinking style and as a vehicle for depicting the outcome of captured processes. The rest of the paper applies this language to a case study taken from the IBM Blueworks literature.

## II. FLOWTHINGS MACHINE (FM)

The FM Model was inspired by the many types of flows running through diverse fields, including information flows, signal flows, and data flows in communication models [7-10]. This model is a diagrammatic schema that uses *flowthings* to represent a range of items, for example, electrical, mechanical, chemical, and thermal signals, circulating blood, prepared and ingested food, shared concepts, pieces of data, activities, and so on. *Flowthings* (hereafter *things*) are defined as what can be *created*, *released*, *transferred*, *processed*, and *received* as stages of a flow machine (hereafter *a machine*), as shown in Figure 1, a generalization of the typical input-process-output model used in many scientific fields (Figure 2).

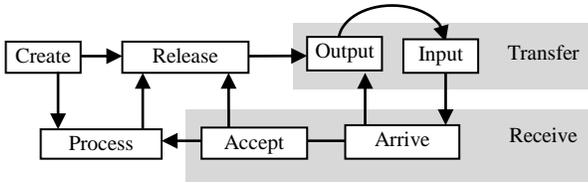

**Figure 1. Flow machine**

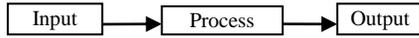

**Figure 2. Input-process-output model**

The stages of the flow machine can be briefly explained as follows:
**Arrive**: A thing reaches a new machine, e.g., data flow to a process
**Accepted**: A thing is approved/not approved to enter a machine, e.g., a datum is not of the right type. If arriving things are always accepted, *Arrive* and *Accept* can be combined as a **Received** stage.
**Processed** (changed): A flowthing goes through some kind of transformation that changes it without creating a new thing, e.g., different number representation.
**Released**: A flowthing is marked as ready to be transferred outside the machine.
**Transferred**: A flowthing is transported somewhere from/to outside the machine.
**Created**: A new thing is born (created) in a machine.

In general, a flow machine is conceived as an abstract machine that receives, processes, creates, releases, and transfers things. The stages in this machine are mutually exclusive (i.e., a thing in the Process stage cannot be in the Create stage or the Release stage at the same time) with respect to atomic flowthings (things that are not constructed from other things). An additional stage of *Storage* can also be added to any machine to represent the storage of things; however, storage is not an exclusive stage because there can be *stored processed* flowthings, *stored created* flowthings, etc.

Flow machines also use the notions of *spheres and subspheres*. These are the network environments and relationships of machines and submachines, e.g., a stomach is a machine in the digestive system *sphere*. Multiple machines can exist in a sphere if needed. A sphere can be a person, an entity (e.g., a company, a customer), a location (a laboratory or waiting room), a communication medium (a channel, a wire). A flow machine is a subsphere that embodies the flow; it itself has no subspheres.

**Example**: Bækgaard and Andersen [11] define an *entity* as a chunk of space-time. An *event* is an entity with defined temporal boundaries, e.g., the Second World War. A *process* is an action composed of identifiable sub-events. The authors [11] propose a basic set of *action* types called interaction primitives: "An interaction primitive can be viewed as a pattern that defines a dynamic relation between two elements. One of the elements performs an action" [11].

The two diagrams shown in Figure 3 represent this type of modeling style. Figure 4 shows the corresponding FM diagrams. In the upper diagram, the book on the shelf flows to the librarian, then to the borrower. The shelf is represented in a barrel shape as way to indicate a storage entity. For simplicity and because only the book flow is shown, the flow machines in *Librarian* and *Borrower* are not enclosed in boxes.

To make the Speaker-Word-Listener diagram (machine) more interesting, we have added *Concepts* and *Understanding* in the lower diagram. The creation of a concept triggers the creation of a word that flows to the listener to trigger the creation of understanding.

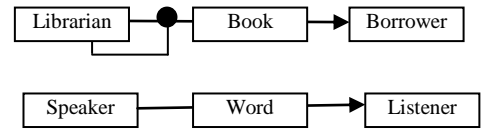

**Figure 3. Some modeling cases (redrawn from** [11]

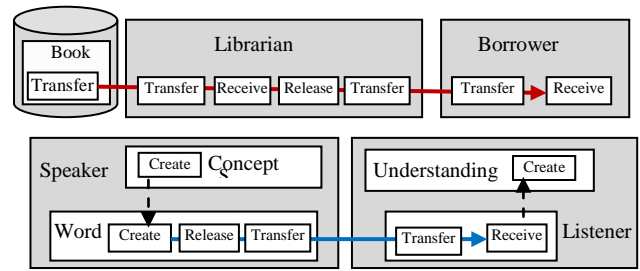

**Figure 4. FM representation, with the lower diagram extended to include flow of *concepts* and *understanding***

### III. CASE STUDY: CALL CENTER COMPANY C

IBM Blueworks (Live) is a flexible, graphical design and visualization tool that enables users to capture business process outlines and create BPMN diagrams.

IBM Blueworks Live does not enforce any process classification framework. How you intend to use the tool for modeling mainly depends on the purpose of your process models. [3]

King et al. [3] give a scenario to demonstrate how to analyze processes and define "solutions" in Blueworks. Their example uses a fictitious company, *Call Center Company C*, that provides call center services to its offices in the US, India, and China. When a large deal is made, Company C cannot serve the huge call volume with its current staff and must hire a significant number of call center representatives (onboarding). The process for hiring new call center employees is manual and not documented.

The onboarding process of call center representatives involves the following players and their functions and requirements**:**

**Recruiter**: identifies potential job candidates and manages communications with them.
**Job candidate**: participates in a job interview, performs a test, and, if he or she accepts a job offer, changes role from job candidate to new hire.
**New hire**: has successfully passed the hiring process but must attend call center training starting work as a call center representative.
**Hiring manager**: performs interviews with job candidates and negotiates contracts, and manages communication with the recruiters. If a new hire's probationary work period is unsuccessful, the hiring manager creates a performance plan.
**Call center manager**: plans the work schedule and runs a probationary review with new employees after 7–10 days of work.
**Human resources (HR) administrator**: helps new employees get started with their first days on the job and also enters employee information into the employee database.

This process for onboarding of new call center employees was performed manually. A team of analysts was established to redefine it by using first IBM Blueworks and then IBM Business Process Manager for implementation of the process to increase process efficiency.

IBM Blueworks is used as the tool to capture process models. The best practice is to first capture the current state process through discovery, then analyze the process information, and, finally, design a future process for IBM BPM implementation" [3].

King et al. [3] start "the blueprinting process" by generating a default process with two *milestones* and one *activity* in the "discovery map view" which involves discovery sessions with the process owner and subject matter experts. This leads to identifying the start and end points of the process, with no focus on exceptional or error conditions. The facilitator listens to the stakeholders and adds *milestones* and *activities*. A milestone is defined as a logical grouping of activities, including,
- a specific *event* occurring or a *condition* met
- a group of activities that represent a *phase* to deliver an output.

In the case study, on the first pass the requirements are separated into the following *logical groups of activities* (milestones):
- Selection of candidates
- Conduct interviews
- Offer negotiation
- Orientation
- Training
- Probationary work review
- Performance plan initiation

After discussions with stakeholders, milestones are converted to activities, and vice versa, e.g., Selection of candidates: *conduct interviews* and *offer negotiation*. Activity is a generic word and can be atomic (task) or compound (process, subprocess, and so on). Then, the seven milestones might be converted to sub-processes.

"Now that you have a sense of the flow of the process, it is time to start adding details to each of the processes and activities" [3]. Next, the following information is gathered: Who supplies the inputs, What are the inputs, What are the processes or activities, What are the outputs, and Who the outputs are for.

Several diagrams are produced in this methodology, as shown in Figures 5, 6, and 7. For example, Figure 7 represents the background check process, involving a criminal background check, a credit check, and a social media background check. The *HR Admin* conducts the background tests and sends the results to the HR Manager for review.

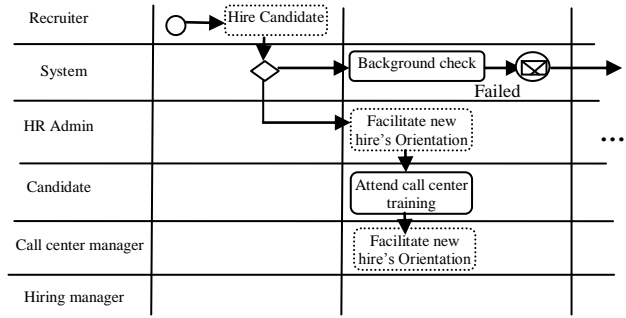

**Figure 5. Call center to-be (redrawn, partial from [3])**

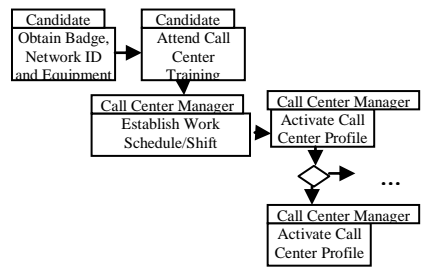

**Figure 6. Playback process flow split (redrawn, partial from [3])**

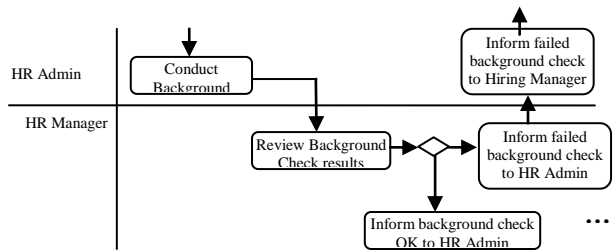

**Figure 7. Background check (redrawn, partial from [3])**

Clearly, it is not possible to present a fair description of the example, given the limitations of a conference paper; however, this description of the case study along with the diagrams provides a general idea of the methodology adopted to capture business process outlines and create a BPMN diagrams.

In the material that follows, we offer an FM-based alternative methodology that could benefit this type of application. To understand the diagrammatic description of Call Center Company C, we have to juxtapose diagrams resulting from the original seven *milestones.* Because juxtaposing several different diagrams is difficult, our understanding of the details of the study case may be imperfect or incomplete; however, these inaccuracies do not detract from our purpose of contrasting two styles of modeling. Any inaccuracies can be redrawn using the FM model notions.

While the Blueworks approach offers a well-proven tool in the field, this utility should not discourage exploration of other lines of thinking about capturing processes. This would benefit the research and also further current proven methodologies such as Blueworks.

## IV. ALTERNATIVE APPROACH

An organization can be described in abstract form as an abstract machine (FM diagram) and realized in reality through its existence over time. *Events, in FM,* are slices of time when sub-machines become "active" to form processes and subprocesses. Figure 8 shows the machine description of *Create job offers* (in black in the online version of the paper). The machine is "activated" by four events: selecting candidates, creating offers, setting deadlines for response, and sending offers. Note that the FM "process" (a stage in the diagram) of an event means that the event *takes its course* after creation. Accordingly, a *process* (in general) is defined as a diagram or subdiagram that incorporates events. Because of its event/time association, a process has a start, a time duration, and an end, and these phases are made obvious in the event descriptions shown in Figure 8.

The Blueworks way of identifying milestones reflects a thinking paradigm that adopts a certain way of generating elements and directs attention to interactions with the world in a way that affects how we link concepts. The Blueworks way of analysis starts as follows:,
- Identify milestones (e.g., in this example, select candidates, conduct interviews, negotiate, …) as disconnected pieces of reality. Then:
- Find the connections among these pieces.

It is a process-less (see the definition of a process given previously) approach that aims at capturing processes in reality.

In this paper, the aim is not to scrutinize the method; rather, the purpose is to explain different philosophies of analysis, as in the case of contrasting the Blueworks approach with the proposed approach in this paper. Use of a certain approach influences the analyst's thinking by "automatically" constraining meanings.

Our proposed approach in the paper aims at modeling reality through processes that *exhibit themselves through flows* (solid and dashed arrows in FM). For example, in Figure 8, we start with Event 1: selecting candidates, and the triggering *leads* us to Event 2 of creating offer. This event *leads* to creating a deadline and starting the flow of an offer to the candidate.

To further contrast Blueworks and our proposed approach, consider the following example from [12].

**Example**: To clarify the milestones-collection methodology of capturing processes in *Call Center Company C*, consider the analogy of a scientist who wants to model the *phosphorus cycle*. In the Blueworks approach, the modeling process is "jump-started" by identifying milestones, as shown in the left diagram of Figure 9. The connections (shown to the right in Figure 9) are found by identifying activities within each milestone.

In contrast, the FM approach begins with the subprocesses of flow of inorganic material in rocks to water, then to ocean, then to sediments, etc., as shown in the steps in Figure 10.

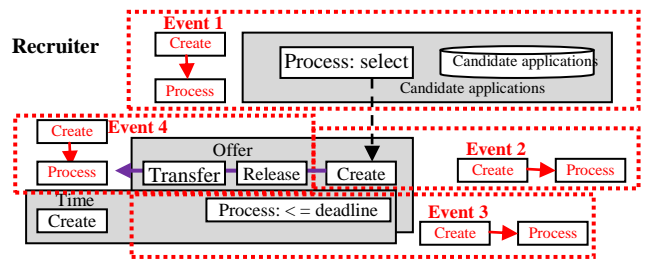

**Figure 8. Illustration of a process formed from four events**

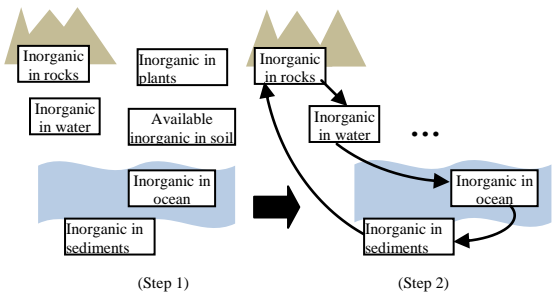

**Figure 9. Blueworks approach begins with identifying milestones**

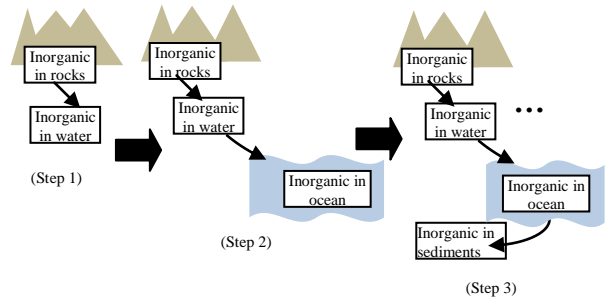

**Figure 10. FM approach to capturing processes**

## V. APPLYING FM TO CALL CENTER COMPANY C

Applying FM to the call center case results in not only a different diagram but also a different developmental method, as explained in the previous section.

Figure 11 shows different phases of building the FM representation. These phases are shown to emphasize the methodology of development. We will focus only on the *final diagram* at the bottom of the figure since it is the culminating representation.

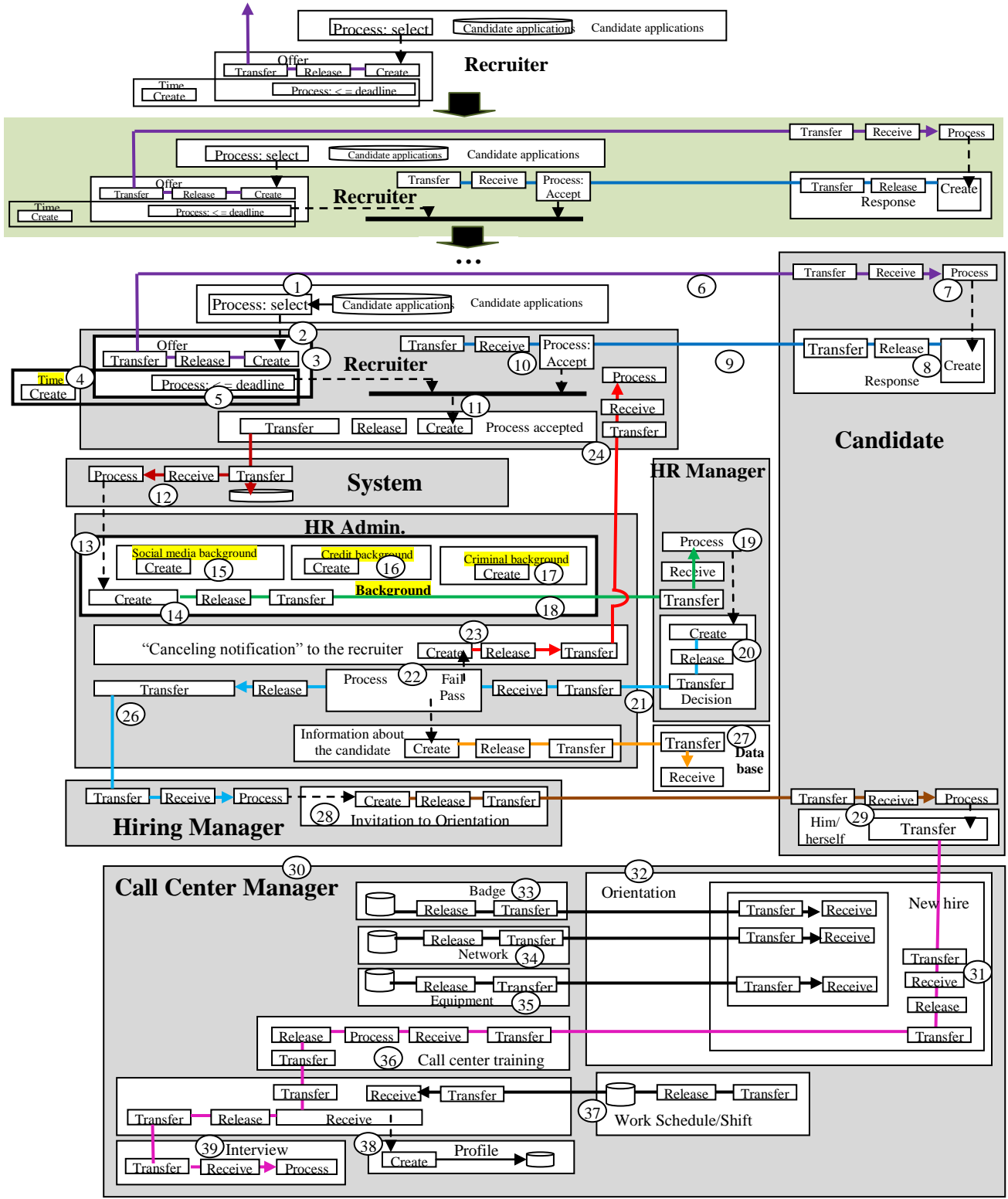

**Figure 11. FM representation of *Call Center Company C***

In Figure 11, the process starts with selection of a candidate (circle 1 in the figure) which triggers (2) the creation of an offer. The offer includes the element of time (4), a thing created universally outside the domain of the recruiter and the processing of time (5) performed by the recruiter. The offer flows to the candidate (6) who processes it (7) to trigger the creation of a response (8) that flows to the recruiter (9). The recruiter processes the response (10); if the candidate has accepted the offer and the deadline period has not expired, then these two conditions trigger (11) the decision to start *the hiring process*. The thick horizontal line indicates that both conditions are realized. The scenario could be modeled in FM as the merging of two flows (of triggering); however, for simplicity, we have used the familiar computer science notation.

This decision to start the hiring process is transferred to the system (12; as described in [3]) and this, in turn, triggers (13) the *HR Admin* to create (14) a background check. The created background check has a structure that is formed from:

Social media background (15)
Credit background (16), and
Criminal background (17).

The background report flows to the HR manager (18), who processes it (19), triggering creation of a hiring decision (20) that flows to the HR manager (21). If the decision is negative (22), a Cancellation notification is sent to the recruiter (23 and 24). If the decision is positive, it is sent to the Hiring manager (26), who registers the candidate's information in the database (27). The Hiring manager also invites (28) the person (29) to an orientation with the Call Center manager (30).

Upon arrival at the call center (31), the person enters an orientation session (32), where he/she is given a badge (33), network account (34), and equipment (35). Then he/she moves to training (36), is given a Work Schedule/Shift (37), creates a profile (38) and finally goes for the interview (39).

## VI. CONCLUSION

This paper has presented a flow-based methodology for thinking about how to capture processes in business process modeling. The proposed methodology is demonstrated through remodeling of a Blueworks case study. The resulting flow-based diagrams reveal a way of thinking in system analysis that is based on flows of things passing through flow machines. Accordingly, an analyst's "thought machine" forms a train of thought that is different from other ways of thinking such as the one reflected in the Blueworks approach. The paper emphasizes this thinking style as a unifying method with potentially diverse applications. The remodeling of the Blueworks case sample seems to point to merits that deserve further development.

## REFERENCES

[1] O. Svatoš, Conceptual process modeling language: Regulative approach. In Proceedings of the 9th Undergraduate and Graduate Students eConf. and 14th Business & Government Executive Meeting on Innovative Cross-border eRegion, Univ. of Maribor, 2007

[2] A. Chapman, Business process modelling, 2016. http://www.businessballs.com/business-process-modelling.htm

[3] J. King, N. Chidambaram, P. Lee, P., Schume, and D. Teran, Process Discovery Best Practices Using IBM Blueworks Live, IBM Redpaper REDP-5111-00, International Technical Support Organization, Sept., 2014.

[4] *IBM,* Process modeling at enterprise scale using BlueworksLive and IBM BPM, WebSphere User Group, IBM Corporation, 2014. http://www.google.com.kw/url?sa=t&rct=j&q=&esrc=s&source=web&cd=1&cad=rja&uact=8&ved=0ahUKEwitm-OXmp_RAhUXN1AKHQEXDroQFgglMAA&url=http%3A%2F%2Fwww.websphereusergroup.co.uk%2Fwug%2Ffiles%2Fpresentations%2F38%2FWUG2014_Blueworks_Live_at_Enterprise_Scale.pdf&usg=AFQjCNEqmhhuqyhb7lTzONwEJA1OKhPQ_A

[5] N. Ward-Dutton, Vendor Insight: IBM breaks new ground with Blueworks Live, Premium Advisory Report, Nov. 2010. https://www.google.com.kw/url?sa=t&rct=j&q=&esrc=s&source=web&cd=2&ved=0ahUKEwjT-IPOnJ_RAhVWeFAKHSb1CukQFggeMAE&url=https%3A%2F%2Fwww.blueworkslive.com%2Fdownload%2Fbpm_vi_IBM_Blueworks_Live_1110.pdf%3FpostId%3Da93c1a523f%26fileItemId%3Da93c1a3981&usg=AFQjCNGgXSERpJmBmb28kwHIU95VWQ6U9Q&bvm=bv.142059868,d.ZWM

[6] R. Peisl, Have fun with collaborative business process modeling in the Cloud with IBM Blueworks Live (BPM), Learnquest Conference, IBM Cloud Technical University, 25–28 October 2016, Madrid, Spain.

[7] S. Al-Fedaghi, "Conceptual modeling in simulation: a representation that assimilates events," Int. J. Adv. Comput. Sci. Appl, Vol. 7, No. 10, 2016, pp. 281-289.

[8] S. Al-Fedaghi, "Toward a philosophy of data for database systems design," Int. J. Database Theory Appl, Vol. 9, No. 10, 2016.

[9] S. Al-Fedaghi, "Diagrammatic modeling language for conceptual design of technical systems: a way to achieve creativity," Int. Rev. Autom. Control, Vol. 9, No. 4, 2016.

[10] S. Al-Fedaghi, 2016. "Flowcharting the meaning of logic formulas," *Int. J. Adv. Res. Artif. Intell. (IJARAI)*, Vol. 5, No. 10, Oct, 2016.

[11] L. Bækgaard, and P. B. Andersen, Using Interaction Scenarios to Model Information Systems, Working Paper I-2008-04, Department of Information and Media Studies, University of Aarhus, 2008. http://www.google.com.kw/url?sa=t&rct=j&q=&esrc=s&source=web&cd=4&cad=rja&uact=8&ved=0ahUKEwj-kfrFop3RAhULIsAKHW5WDusQFggtMAM&url=http%3A%2F%2Fpure.au.dk%2Fportal%2Ffiles%2F2570%2FI_2008_04.PDF&usg=AFQjCNFqJFyjBumdKkIu5ZR5YHiCEN90Eg

[12] Science Learning Hub, The phosphorus cycle, University of Waikato, 2007, http://sciencelearn.org.nz/Contexts/Soil-Farming-and-Science/Sci-Media/Images/The-phosphorus-cycle

AUTHORS PROFILE

Dr. Sabah Al-Fedaghi holds an MS and a PhD in computer science from the Department of Electrical Engineering and Computer Science, Northwestern University, Evanston, Illinois, and a BS in Engineering Science from Arizona State University, Tempe. He has published two books and more than 260 papers in journals and conferences on software engineering, database systems, information systems, computer/information privacy, security and assurance, information warfare, and conceptual modeling. He is an associate professor in the Computer Engineering Department, Kuwait University. He previously worked as a programmer at the Kuwait Oil Company and headed the Electrical and Computer Engineering Department (1991–1994) and the Computer Engineering Department (2000–2007).